\documentclass{aa}  

\usepackage{graphicx}
\usepackage{txfonts}
\def\Teff{\ensuremath{T_\mathrm{eff}}}
\def\logl{\ensuremath{\log (L/L_{\sun})}}
\def\ubv{\ensuremath{UB\emph{V}}}
\def\bvri{\ensuremath{B\emph{V}(RI)_{C}}}
\def\bvi{\ensuremath{B\emph{V}I_{C}}}
\def\uvbyb{\ensuremath{uvby\beta}}

\begin{document}

   \title{Towards a photometric metallicity scale for open clusters}


   \author{M. Netopil
          \inst{1}
          \and
          E. Paunzen\inst{2,3}}

   \institute{Institute for Astrophysics, University of Vienna,
              T\"{u}rkenschanzstra{\ss}e 17, A-1180 Vienna \\
              \email{martin.netopil@univie.ac.at}
         \and
             Department of Theoretical Physics and Astrophysics, Masaryk University, Kotl\'a\v{r}sk\'a 2, 611 37 Brno, Czech Republic \\
             \email{epaunzen@physics.muni.cz}
         \and 
              Rozhen National Astronomical Observatory, Institute of Astronomy of
              the Bulgarian Academy of Sciences, P.O. Box 136, BG-4700 Smolyan, Bulgaria}   
             

   \date{Received <date>; accepted <date>}

 
  \abstract
   {Open clusters are a useful tool when investigating several topics connected with stellar evolution; for example the age or distance can be more accurately determined than for field stars. However, one important parameter, the metallicity, is only known for a marginal percentage of open clusters.} 
   {We aim at a consistent set of parameters for the open clusters investigated in our photometric $\Delta a$ survey of chemically peculiar stars. Special attention is paid to expanding our knowledge of cluster metallicities and verifying their scale.}
   {Making use of a previously developed method based on normalised evolutionary grids and photometric data, the distance, age, reddening, and metallicity of open clusters were derived. To transform photometric measurements into effective temperatures to use as input for our method, a set of temperature calibrations for the most commonly used colour indices and photometric systems was compiled.}
   {We analysed 58 open clusters in total. Our derived metallicity values were in excellent agreement with about 30 spectroscopically studied targets. The mean value of the absolute deviations was found to be 0.03\,dex, with no noticeable offset or gradient. The method was also applied using recent evolutionary models based on the currently accepted lower solar abundance value Z\,$\sim$\,0.014. No significant differences were found compared to grids using the former adopted solar value Z\,=\,0.02. Furthermore, some divergent photometric datasets were identified and discussed.}
   {The method provides an accurate way of obtaining properly scaled metallicity values for open clusters. In light of present and future homogeneous photometric sky surveys, the sample of stellar clusters can be extended to the outskirts of the Milky Way, where spectroscopic studies are almost impossible. This will help for determining galactic metallicity gradients in more detail.}


   \maketitle
%

\section{Introduction}

Open clusters are excellent astrophysical laboratories for almost all important
processes connected with stellar formation and evolution. These include the mechanisms
of diffusion, rotation, mass loss, pulsation, and accretion. One can study not only astrophysical processes, but also the connection between various star classes (e.g. variable stars) and the local metallicity or age. Star clusters are also immensely important tracers of the
inner and outer galactic spiral arms.

One important characteristic of stellar formation, the metallicity, is poorly known for galactic open clusters. In the updated catalogue by \citet{dias02} [Version 3.3.\footnote{http://www.astro.iag.usp.br/$\sim$wilton/}], this parameter is listed for about 200 clusters, fewer than for 10\,\% of the currenty known population. These values were derived from a wide variety of applied methods, ranging from metallicity-dependent photometric indices and isochrone fitting, to low- and high-resolution spectroscopy. The usual convention for expressing metallicity is the logarithmic term [Fe/H], which represents the abundance ratio of iron to hydrogen in the stellar atmosphere.

\citet{carrera11} have compiled an almost complete list of clusters studied with medium- or high-resolution (R\,\gid\,15\,000) spectroscopy. By also including recent literature, the spectroscopically investigated sample comprises about 100 targets so far. Most targets were studied only once, and the mean cluster metallicities were on average based on a maximum of three stars. These are generally red giants since they are the brightest cluster objects. However, \citet{santos12} find differences for white dwarfs and giant stars, mainly because of the use of improper line lists for evolved objects. 

Some photometric systems also provide metallicity estimates (e.g. the Str\"omgren, Johnson, or DDO systems), but these also suffer from temperature range restrictions. In the work by \citet{phn10}, photometric results were compiled for 188 open clusters in total. This list also shows that the
majority of targets were investigated only once with data for only a few stars, and large differences were noticed when comparing
various sources.

Studies based on colour-magnitude diagrams and isochrone fitting usually neglect the metallicity parameter, setting it for the sake of simplicity to a solar value for the targets \citep[e.g.][]{Khar05,subra10,glu13}. \citet{poe10}, on the other hand, present a method that offers a valuable approach to investigating metallicities using evolutionary models and photometric data of all main-sequence stars in a cluster. Nevertheless, it is necessary to investigate a larger sample of objects to verify whether this method provides properly scaled results.

\section{Method and target selection}
\label{method}
To investigate open cluster parameters in a consistent way, we applied the method by \citet{poe10}. They calculated differential evolutionary tracks (normalised to the zero age main-sequence, ZAMS) for a variety of metallicity/age combinations. These have to be compared with the observed Hertzsprung-Russell diagram (HRD) of open-cluster main-sequence stars. Since spectroscopically determined effective temperatures are not	readily available, photometric data have to be transformed to the $\Teff$/$\logl$ plane. Using an iterative procedure, the cluster parameters are altered until the best final solution for all parameters is found. It is beneficial when some of the input parameters (the age, reddening, and distance) can be restricted (e.g. available Hipparcos parallax, reddening deduced from photometry). However, incorrectly adopted starting values can be recognised during this procedure. Where targets had distances derived from Hipparcos data \citep{leeuwen09}, they were utilised as the initial parameter. Otherwise, we made use of the literature compilation of cluster parameters by \citet{pn06}. This had been updated with recent investigations to obtain proper mean starting values for the age, distance, and reddening of our programme clusters. 

We followed the procedure by \citet{poe10}, but made use of a broader selection of photometric data by incorporating among others near-infrared (NIR) 2MASS measurements \citep{strut06} for some targets. Whenever possible, we derived individual reddening values for the cluster stars. For most targets, $\ubv$ or Geneva photometry are available, which allowed us to determine the colour excess for O/B type stars by means of the \textit{Q} method \citep{J58} and X/Y parameters \citep{cram82}, respectively. Even cooler type stars could be de-reddened with $\uvbyb$ data and the appropriate calibrations \citep[e.g.][]{na93}.   

The target open clusters originate in the photometric $\Delta a$ survey \citep[e.g.][]{net07}, which is dedicated to the detection of chemically peculiar (CP) stars. So far, about 80 open clusters have been covered. A homogeneous set of cluster parameters is essential to investigating possible dependencies between the occurrence of CP objects and metallicity or age, for example. However, very young open clusters cannot be treated safely with the applied method owing to the too inconspicuous deviation from the ZAMS and to the restriction of the grids to $\log t \geq 7.2$. Open clusters with strong differential reddening, but without sufficient photometry to determine individual reddening values, were excluded as well. Nevertheless, we analysed a significant number of 58 open clusters (listed in Table \ref{results}) with the differential grid (DG) method. 

This sample incorporates about 30 targets with available spectroscopic metallicity determinations. A comparison with our results thus allowed us to verify that the DG method could provide a proper metallicity scale for open clusters (see Sect. \ref{speccomp}).

\section{Defining the temperature scale}
\label{teffscale}
To apply the DG method, it is important to define a proper temperature scale for an accurate transformation of photometric data to effective temperatures. In contrast to \citet{poe10}, who mostly used $(B-V)$ data as an input, we utilised a broader selection of temperature dependent colour indices and photometric systems. This guaranteed an increase in the accuracy of the temperature, but also allowed us to recognise erroneous photometric data. As did \citet{poe10}, we adopted the temperature calibration by \citet{Alonso96} for the colour index $(B-V)$, as well as their relation for $(V-R)_{J}$. Since the colour index is defined for the original Johnson system, and CCD studies are generally performed in the Cousins system, the following transformation by \citet{Bess83} needs to be applied: 
$$(V-R)_{C} = 0.715(V-R)_{J} - 0.02.$$ 
To cover a greater wavelength range, we considered the almost metallicity-independent colour indices $(V-I)$ and $(V-K)$, respectively \citep[see][]{Alonso96}. This reference provided temperature calibrations for stars cooler than spectral type F0 ($\sim$\,8\,000\,K). We therefore adopted the colour-temperature relations by \citet{Bess98} for $(V-I)_{C}$ and by \citet{bened98} for $(V-K)_{J}$, both valid up to about 10\,000\,K. Once again, the latter colour index is in the Johnson system, but since we incorporate 2MASS $K_{s}$ data for some open clusters, these need the correction given by \citet{carpenter01}. To extend the $(B-V)$ temperature calibration by \citet{Alonso96} to stars earlier than spectral type F0, we used the results by \citet{FLO96}. This author provided a compilation of ``fundamental'' temperatures \citep[for example from][]{code76} for numerous stars of different luminosity classes and derived temperatures based on $(B-V)$ and a calibration for the bolometric correction as a function of colour/temperature. We adopted his list, but used only stars of luminosity class IV/V with available photometric data in other passbands taken from the General Catalogue of Photometric Data (GCPD)\footnote{http://obswww.unige.ch/gcpd/gcpd.html}.

\begin{figure}
\centering
\includegraphics[width=88mm]{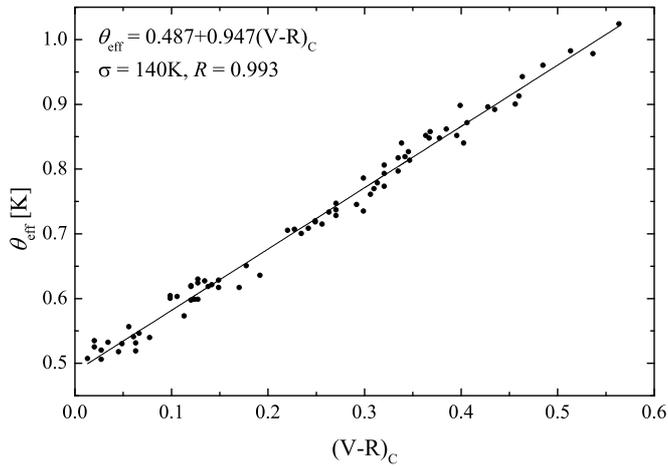}
\caption{Empirical temperature calibration for the colour index $(V-R)_{C}$.}
\label{vrcal}
\end{figure}

Based on a sample of 82 objects that were almost free from reddening, we defined an extension to the temperature calibration by \citet{Alonso96} for $(V-R)_{C}$, which is valid for the range between 5\,000 to 10\,000\,K, with a mean standard deviation $\sigma = 140$\,K and a correlation coefficient $R = 0.993$ (see Fig. \ref{vrcal} and equation below):

$$\theta_\mathrm{eff} = 0.487(3) + 0.947(13) (V-R)_{C}.$$

The errors of the last significant digits are given in parentheses. Noteworthy is that the colour index is already in the Cousins system, and $\theta_\mathrm{eff}$ is defined as 5040/$\Teff$ in order to avoid a higher order polynomial fit. The fit is in excellent agreement with \citet{Alonso96} in the overlapping temperature range, with differences of less than 50\,K.

\begin{figure}
\centering
\includegraphics[width=88mm]{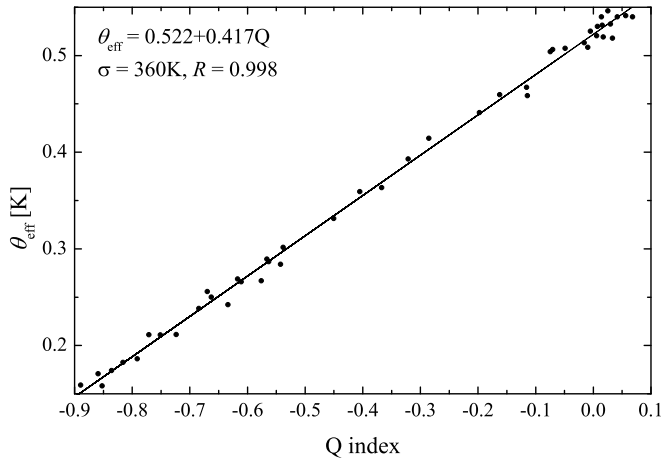}
\caption{Empirical temperature calibration for the reddening free $\ubv$ $Q$-index.}
\label{qindex}
\end{figure}

\begin{figure}
\centering
\includegraphics[width=88mm]{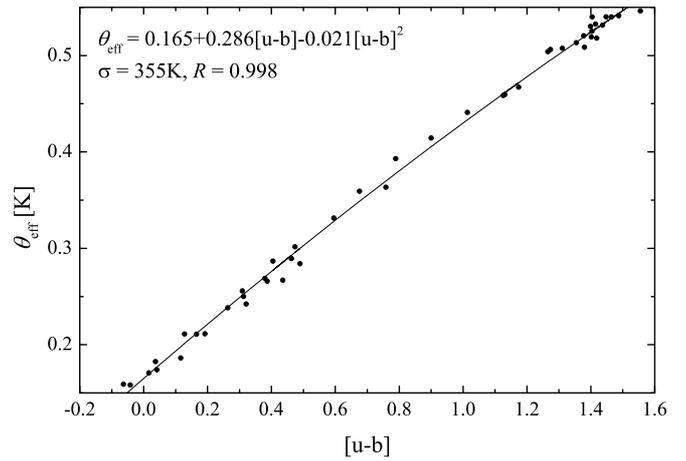}
\caption{Empirical temperature calibration for the reddening free $[u-b]$ Str\"omgren index.}
\label{ubindex}
\end{figure}

For objects hotter than 9\,500\,K it is more efficient to use reddening-free indices, such as Johnson $Q$ or Str\"omgren $[u-b]$. We made use of the refined definition $Q= (U-B) - 0.71(B-V)$ by \citet{Bess98}. For completeness, the $[u-b]$ index is defined as
$[u-b]=[c_{1}]+2[m_{1}]$ with $[c_{1}]=c_{1}-0.2(b-y)$ and $[m_{1}]=m_{1}+0.32(b-y)$ according to \citet{crawmand76}. As for the temperature relation presented above, we made use of stars listed by \citet{FLO96}, but selected only the hotter ones (\ga\,9\,500\,K). This sample was extended with temperatures for O/B type stars presented in the papers by \citet{morel08}, \citet{prz08}, \citet{simon10}, and \citet{nieva11}. Furthermore, we included results by \citet{lefever10}, however only their ``well-studied'' objects were adopted (see the reference for details). No significant offsets or trends (within the respective errors) were found between common stars of the different studies, we therefore adopted mean values for the subsequent analysis. The compiled list of stars was checked for luminosity class (IV/V) and variability, resulting in the exclusion of several $\beta$~Cephei type stars or slowly pulsating B-type objects. In total, 46 objects define our final list of fundamental temperatures. Their relation to the Johnson $Q$ and Str\"omgren $[u-b]$ indices are presented in Figs. \ref{qindex} and \ref{ubindex}, respectively, with mean standard deviations of about 350\,K and correlation coefficients $R = 0.998$ for both empirical calibrations:

$$\theta_\mathrm{eff} = 0.522(2)+0.417(4)Q$$ 
$$\theta_\mathrm{eff} = 0.165(3)+0.286(10)[u-b]-0.021(7)[u-b]^2.$$

\begin{table}
\caption{Adopted temperature calibrations.} 
\label{adopttemp} 
\centering 
\begin{tabular}{l l} 
\hline\hline 
Colour / System & calibration \\ 
\hline 
$Q$ index ($\ubv$)& this study \\
$(B-V)$ & \citet{Alonso96} / \citet{FLO96}\tablefootmark{a} \\
$(V-R)$ & \citet{Alonso96} / this study \tablefootmark{a}\\
$(V-I)$ & \citet{Bess98} \\
$(V-K)$ & \citet{bened98} \\
$Geneva$ & \citet{kunz97} \\
$\uvbyb$ & \citet{na93} \\
$(b-y)$ & \citet{na93} \\
$[u-b]$ & this study \\
\hline 
\end{tabular}
\tablefoot{
\tablefoottext{a}{For stars cooler/hotter than $\sim$\,8\,000\,K as described in the text.}
}
\end{table}

In addition to the temperature calibrations already discussed, we applied the widely accepted routines by \citet{kunz97} to the Geneva photometry, and the UVBYBETA code developed by \citet{na93} to the $\uvbyb$ data. These authors also provided calibrations in terms of $(b-y)$ and $[u-b]$, the latter based on only 14 objects. A comparison with our $[u-b]$ scale shows that both are in reasonable agreement, although the temperatures by \citet{na93} around 30\,000\,K are underestimated by about 2\,\%.

\section{The photometric data and member selection}
\label{photscale}

The WEBDA\footnote{http://webda.physics.muni.cz} database is probably the most valuable source available for open cluster data (photometry and auxiliary information such as membership probabilities). With the implemented tools, the database also allows for an initial comparison of different datasets. We retrieved almost all available data for our programme clusters from it. Photographic measurements were not directly used for the presented study owing to their high associated uncertainties. However, they were used to verify the other photometry. 

To identify possible erroneous datasets, a comparison of all available measurements was conducted. Using the equations by \citet{harmanec01}, we also transformed the photometry between the $\ubv$, $\uvbyb$, and Geneva systems, in order to check the individual zero points. However, compared to some CCD studies discussed below, the photoelectric studies showed hardly any significant offsets or gradients. Therefore, we used photoelectric data instead of CCD measurements, when the photoelectric data are sufficiently complete down to the lower mass stars.  

\citet{pau02} noticed that the CCD $V$ measurements for NGC~6451 by \citet{pcb98} were too bright compared to other literature results. We determined an offset of 0.97\,mag using data by \citet{kjeld91}. However, the provided $(V-I)_{C}$ colours by \citet{pcb98} were not influenced. This was verified by comparing the resulting $I$ magnitudes to those from the DENIS survey\footnote{VizieR online data catalogue: II/263.}. In a follow-up paper, \citet{pau03} investigated this cluster and NGC~6192 with $uvby$ filters. The last object has been studied well in \ubv, so that a comparison using the transformations by \citet{harmanec01} revealed that the $V$ magnitudes are on the correct scale. However, especially $(u-b)$ was inaccurately standardised, probably owing to the use of only three spectrophotometric standard stars. This led to differing temperatures based on the routine by \citet{na93} and the $[u-b]$ calibration defined in our study. We therefore excluded the datasets for both clusters. 


\citet{kjeld91} provided $\ubv$ CCD photometry for 13 open clusters in total, with six of them among our programme targets. They mentioned that their transformation to standard $(U-B)$ colours is highly inaccurate. We therefore have not incorporated this colour index from the latter reference. The remaining colours appear correctly transformed and were considered for our analysis.

Recently, \citet{glu13} have investigated the cluster NGC~7296 by means of $\bvri$ photometry. They notice some offsets in the data by \citet{net05}, by comparing these to their measurements and the ones of the AAVSO Photometric All-Sky Survey (APASS\footnote{http://www.aavso.org/apass}). Therefore, instead of correcting the previously available ones, we adopted only the measurements by \citet{glu13}. They offered a much better cluster coverage and additional colour information.
 
For the open cluster Ruprecht~130, the only available photometric studies are the ones by \citet{piatti00} by means of $\bvi$ and \citet{pau06}. The latter reference used the former to calibrate their $\Delta a$ photometry. Using the Guide Star Catalog for an initial check of the photometry, an offset similar to the one found for NGC~6451 can be seen. According to the observation logs, one can infer that the data presented by \citet{pcb98, piatti00} were obtained during the same observing run. We therefore applied the same offset to the $V$ magnitudes as determined for NGC~6451 (+0.97\,mag). Again, the resulting $I$ magnitudes were compared with the Denis survey, showing excellent agreement ($\sim$\,0.01\,mag). Furthermore, some cluster stars were covered by the APASS survey. The stellar magnitudes corresponded well to the corrected $V$ measurements and the $(B-V)$ colours from \citet{piatti00}. To obtain the additional colour index $(V-K)$ for the stars in Ruprecht~130 (see Sect. \ref{teffscale}), we queried the 2MASS catalogue. However, this cluster is very faint, and data with good (``A'' quality flag) photometry are only available for the brightest objects. We therefore used the DR8plus release of the UKIDSS Galactic Plane Survey \citep[GPS,][]{lucas08} for an extension to fainter stars. However, in this area we noticed the following offset and colour dependencies between the UKIDSS and 2MASS photometry. This was based on about 200 common stars with the highest quality flags in both surveys: 

$$J_{2M} = J_{UK} + 0.068(39)$$
$$(J-H)_{2M} = 0.133(6) + 1.063(7)(J-H)_{UK}$$
$$(J-K)_{2M} = 0.092(10) + 1.025(8)(J-K)_{UK}.$$

The UKIDSS GPS survey uses 2MASS in combination with extinction values from the dust maps by \citet{schlegel98} to define the photometric zero-points \citep[see][for details]{lucas08}. The reddening value provided by these dust maps is several magnitudes higher than the actual value for the cluster (see Sect. \ref{resdisc}). Thus, the UKIDSS photometry is probably influenced in this area. The corrected UKIDSS data were merged with 2MASS photometry to cover the cluster from the brightest to faintest objects. We selected only the most accurate measurements using the photometric quality flags from both datasets.

We note that the aforementioned APASS survey is also still in progress. It currently provides aperture photometry (17\arcsec diameter), resulting in crucial blends in denser areas. Taking this limitation into account by selecting isolated stars, we incorporated APASS data for some closer poorly covered clusters such as Trumpler~10. 

Only some photoelectric $\ubv$ measurements by \citet{mohan84} are available for the brightest stars in the open cluster King~21. We therefore used this dataset, together with APASS photometry, to recalibrate the $\Delta a$ photometry by \citet{net07}, in order to obtain at least $V$ and $(B-V)$ for the fainter stars. The complete available data set of the APASS survey cannot be safely used for the reason described previously. However, because King~21 is a young cluster and we are not able to derive individual reddening values with the resulting data, a restriction to temperatures lower than 10\,000\,K was made to avoid erroneous results. Furthermore, we applied calibrated $\Delta a$ photometry for the analysis of NGC~6830, another poorly studied cluster.

Finally, if several studies in the same colour were available for individual cluster stars, their mean values were adopted for the analysis. One exception was the cluster NGC~6705, which was covered by a $\bvi$ standard field sequence \citep{stet00}. We therefore used only the latter work, combined with $(U-B)$ data from \citet{sung99}.

As mentioned in Sect. \ref{method}, the DG method makes use of main-sequence stars. The most evolved objects (red giants) can be easily recognised in the various colour-magnitude diagrams (CMD), and the remainder of non-main-sequence stars are noticeable during final analysis by their large deviation from the differential grids in luminosity and temperature.

The cluster membership of the individual objects was determined by means of all available CMD and colour-colour diagram combinations. Furthermore, the kinematic membership probabilities and spectral types listed in WEBDA were consulted. In most cases, temperatures were determined using several calibrations. If significant differences are found among individual results, the objects were excluded in order to obtain a sample of only the most probable cluster stars with accurate photometry for further analysis .  

\citet{maicat86} proposed that there are two clusters at different distances in the direction of NGC 2451. This has been confirmed by several studies \citep[][]{hunsch04}. Since the mean proper motions differ significantly \citep[see e.g.][]{Khar05}, the respective members can be easily distinguished. We therefore used Tycho-2 proper motions and additional literature references \citep[e.g.][]{platais01} to extract 30 and 29 members for NGC~2451 A and B, respectively.

\section{Results and discussion}
\label{resdisc}
Using the temperature calibrations summarised in Table \ref{adopttemp}, the compiled photometric data for the programme open clusters were transformed into effective temperatures. Whenever possible, individual reddening values for the cluster stars were derived. We used the reddening-ratios listed by \citet{Bess98} to transform $E(B-V)$ to other colours in the Johnson-Cousins system, and Str\"omgren and Geneva reddening values were transformed as follows \citep[see e.g.][]{net08}: 
$E(B - V ) = 1.43E(b - y) = 0.84 E[B - V ] = 1.14 E(B2 - V 1) = 0.83 E(B2 - G)$.
If the available data did not allow for determining individual reddening values, a mean cluster reddening value was calculated from compiled literature results and used as the starting value. This was also true for the remaining parameters: age and distance. 

Finally, averaged temperatures were used to derive the bolometric corrections \citep[][]{FLO96} needed to obtain luminosity. In an iterative procedure, the input cluster parameters were altered until the best fit with the grids by \citet{poe10} was found (lowest $\sigma$ over the complete luminosity range). A consistency check of the derived parameter set was performed by fitting Geneva \citep{ls01} and Padova \citep{marigo08} isochrones to the CMDs. It was possible to apply the DG method to 58 clusters in total. For the remaining clusters, there are too few usable photometric data available. However, the majority of the clusters that we did not investigate were simply too young for this method.  

The final results are shown in Table \ref{results} and Figs. \ref{plot1} to \ref{plot10} (the figures are available only in the online version). The DG method provides the overall metallicity (Z), a parameter that is rarely known for open clusters or even single stars. To allow a comparison with other (e.g. spectroscopic) studies, these were transformed into the more common iron abundance ratio [Fe/H] as given by \citet{poe10}.

We have to note that it is difficult to list a representative error for the fits. Although the standard deviation of Z (as listed in Table \ref{results}) can be used as an estimate, also the number of objects, the coverage down to solar luminosity, and the accuracy and number of available photometric datasets have to be considered. A good example is the poorly populated old open cluster NGC~1901. It only has 11 main-sequence member stars that are usable for the metallicity determination. We derived a value of Z=0.019(2); however, the given error is very probably underestimated, because the mean standard deviation of all investigated clusters is twice as large.

From our sample there are ten targets in common with the investigation by \citet{poe10}, allowing a direct comparison of the results. There was little deviation between the derived metallicities from the two studies. However, we noticed three clusters whose iron abundances deviate by more than 0.10\,dex. For two of them (NGC~2516 and NGC~7092) the discrepancy could be explained by the applied distances. Especially for NGC~2516, the difference between the true distance moduli used amounts to $-$\,0.27\,mag (former minus present study). Our distance scale is in line with the comprehensive investigation by \citet{an07}, who used main-sequence fitting with empirically calibrated isochrones. The results for the third strongly deviating cluster (Melotte~20) were probably affected by incorrect effective temperatures derived by \citet{poe10}. We noticed that the temperatures of their most luminous objects were underestimated by about 3000\,K. This demonstrates the benefit of incorporating several photometric systems, the use of reddening free indices for hotter stars, and individually determined reddening values compared to adopting a single colour index and a uniform reddening value for all objects. The effective temperatures for Melotte~20 stars derived by us, using $\ubv$, $\uvbyb$, and Geneva photometry, are consistent with a mean standard deviation of 150\,K. Owing to the lower temperatures for the luminous stars, \citet{poe10} also obtained a much older age for Melotte~20 compared to our analysis and those quoted in the literature so far. No inconsistencies for the other open clusters in common were noticed.

\begin{figure}
\centering
\includegraphics[width=88mm]{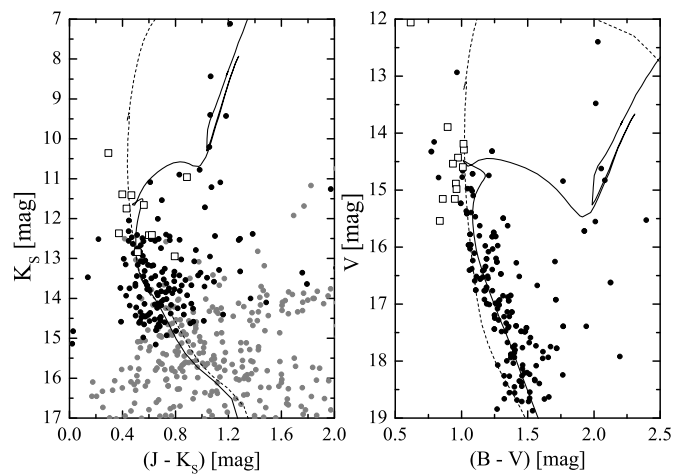}
\caption{The colour-magnitude diagrams for Ruprecht~130. Black dots in the NIR CMD represent the stars observed in the visual by \citet{piatti00}. Grey dots are additional 2MASS/UKIDSS objects within a 2\,\arcmin radius. Open squares indicate the position of stars near the blue hook and probable blue stragglers, which are noticeable in the visual, with available good quality NIR data. The solid lines show the isochrones for our results (560\,Myr), and the dashed lines 50\,Myr isochrones with the parameters listed by \citet{piatti00}.
}
\label{rup130cmd}
\end{figure}

The largest deviation between starting values and final results was found for Ruprecht~130, an open cluster already discussed in Section \ref{photscale}. The derived distance of 2000\,pc agrees to within 10\,\% of the previous values determined by \citet{piatti00} and \citet{pau06}, although these values are based on $\sim$\,1\,mag brighter $V$ magnitudes and a 0.20\,mag higher reddening value. However, the most outstanding difference was noticed for the age. Our analysis resulted in a cluster age of 560\,Myr, whereas the previous studies obtained ages of 50 and 80\,Myr. Both studies performed isochrone fitting, however \citet{piatti00} also matched integrated spectra to available templates. Figure \ref{rup130cmd} shows the NIR and magnitude corrected visual CMD overlayed with solar composition isochrones by \citet{marigo08}. The apparent distance modulus given by \citet{piatti00} was corrected to account for the derived magnitude offset. 
For the NIR isochrones, the parameters were transformed using the relations $E(J-K_{S}) = 0.488E(B-V)$ by \citet{bon04} and $A_{K_{S}} = 0.67E(J-K_{S})$ by \citet{dutra02}.

It is obvious that in the visual region the previous parameters do not provide a proper fit for the main-sequence at all. Furthermore, the age is set by the brightest and bluest objects. In contrast, our determined parameters cover the probable red giant branch, noticeable especially in the NIR diagram. We would like to draw attention to the fact that the results of the DG method are based solely on the main-sequence. In light of the older age, the aforementioned brightest and bluest objects can be assigned to the blue hook or to the group of blue stragglers.

\begin{table*}
\caption{The results for the programme targets.} 
\label{results} 
\centering 
\tiny
\begin{tabular}{l c r l c l l l} 
\hline\hline 
Cluster & $\log t$ & $(m-M)_{0}$ & $E(B-V)$\tablefootmark{a} & Z & [Fe/H]$_{DG}$ & [Fe/H]$_{spec}$\tablefootmark{b} & phot. systems\tablefootmark{c} \\ 
\hline 
Berkeley 11 & 7.80 & 12.20 & 0.96(6) & 0.020(5) & +0.01(14) &   & UBV$_{pe}$, UBVRI$_{ccd}$ \\
Collinder 140 & 7.55 & 8.05 & 0.04(4) & 0.021(3) & +0.04(7) &   & UBV$_{pe}$, Str$_{pe}$, G \\
Collinder 272 & 7.30 & 11.75 & 0.43(3) & 0.021(4) & +0.06(9) &   & UBVRI$_{ccd}$ \\
IC 2391 & 7.65 & 5.80 & 0.01(1) & 0.018(3) & $-$0.03(7) & $-$0.02(1)/2 & UBV$_{pe}$, Str$_{pe}$, G \\
IC 2602 & 7.85 & 5.85 & 0.03(2) & 0.019(3) & +0.00(8) & $-$0.03(4)/2 & UBV$_{pe}$, Str$_{pe}$, G \\
IC 4665 & 7.70 & 7.70 & 0.19(3) & 0.020(2) & +0.02(6) & $-$0.03(4)/1 & UBV$_{pe}$, Str$_{pe}$, G \\
IC 4725 & 7.85 & 9.00 & 0.45(5) & 0.020(3) & +0.03(8) & +0.02(2)/1 & UBV$_{pe}$, Str$_{pe}$ \\
King 21 & 7.30 & 12.40 & 0.86(6) & 0.021(6) & +0.06(14) &   & UBV$_{pe}$, $\Delta a$, 2M \\
Lynga 1 & 8.10 & 11.40 & 0.45(4) & 0.021(4) & +0.04(10) &   & UBV$_{pe}$, UBVRI$_{ccd}$ \\
Melotte 20 & 7.75 & 6.15 & 0.09(3) & 0.021(5) & +0.05(13) & +0.09(20)/2 & UBV$_{pe}$, Str$_{pe}$, G \\
Melotte 22 & 8.05 & 5.50 & 0.04(2) & 0.016(2) & $-$0.08(7) & +0.02(4)/5 & UBV$_{pe}$, Str$_{pe}$, G \\
Melotte 105 & 8.60 & 11.85 & 0.44(4) & 0.022(5) & +0.08(12) &   & UBVI$_{ccd}$, Str$_{ccd}$ \\
Melotte 111 & 8.85 & 4.70 & 0.00(1) & 0.018(2) & $-$0.04(5) & +0.01(8)/2 & UBV$_{pe}$, Str$_{pe}$, G \\
NGC 1039 & 8.25 & 8.35 & 0.08(2) & 0.021(4) & +0.05(10) & +0.07(4)/1 & UBV$_{pe}$, BV$_{ccd}$, Str$_{pe}$, G \\
NGC 1662 & 8.60 & 8.05 & 0.32(3) & 0.021(3) & +0.05(8) &   & UBV$_{pe}$, Str$_{pe}$, G \\
NGC 1901 & 8.90 & 8.20 & 0.03(2) & 0.019(2) & +0.01(5) & $-$0.08(1)/1 & UBV$_{pe}$, UBV$_{ccd}$, Str$_{pe}$ \\
NGC 2099 & 8.65 & 10.55 & 0.30(1) & 0.019(2) & +0.00(4) & +0.01(5)/1 & UBV$_{pe}$ \\
NGC 2232 & 7.70 & 7.65 & 0.03(2) & 0.025(6) & +0.14(12) & +0.22(9)/1 & UBV$_{pe}$, Str$_{pe}$, G \\
NGC 2287 & 8.40 & 9.10 & 0.02(2) & 0.014(3) & $-$0.16(9) & $-$0.23(2)/1 & UBV$_{pe}$, Str$_{pe}$, G \\
NGC 2343 & 8.05 & 9.90 & 0.20(2) & 0.021(6) & +0.05(14) &   & UBV$_{pe}$, by$_{ccd}$ \\
NGC 2422 & 8.15 & 8.50 & 0.09(2) & 0.023(4) & +0.09(8) &   & UBV$_{pe}$, UBVI$_{ccd}$, Str$_{pe}$, G \\
NGC 2423 & 9.00 & 9.80 & 0.10(1) & 0.024(5) & +0.11(10) & +0.14(6)/1 & Str$_{pe}$, G, APASS \\
NGC 2447 & 8.75 & 10.13 & 0.01(1) & 0.020(3) & +0.03(7) & $-$0.03(6)/4 & Str$_{pe}$, G \\
NGC 2451A & 7.70 & 6.35 & 0.01(1) & 0.017(4) & $-$0.06(11) &   & UBV$_{pe}$, BV$_{ccd}$, Str$_{pe}$, G \\
NGC 2451B & 7.70 & 7.95 & 0.10(3) & 0.020(3) & +0.02(9) &   & UBV$_{pe}$, Str$_{pe}$, G \\
NGC 2489 & 8.75 & 11.30 & 0.31 & 0.022(4) & +0.06(10) &   & UBVI$_{ccd}$ \\
NGC 2516 & 8.25 & 8.05 & 0.11(3) & 0.020(5) & +0.03(12) & +0.01(7)/1 & UBV$_{pe}$, Str$_{pe}$, G \\
NGC 2546 & 8.15 & 9.75 & 0.14(3) & 0.020(4) & +0.01(9) &   & UBV$_{pe}$, APASS, 2M \\
NGC 2567 & 8.45 & 11.05 & 0.13(3) & 0.020(4) & +0.02(10) & +0.00(5)/1 & UBV$_{pe}$ \\
NGC 2632 & 8.80 & 6.30 & 0.01(1) & 0.027(5) & +0.17(9) & +0.14(10)/4 & UBV$_{pe}$, Str$_{pe}$, G \\
NGC 2658 & 8.50 & 13.05 & 0.36(3) & 0.020(5) & +0.02(12) &   & UBVRI$_{ccd}$ \\
NGC 3114 & 8.20 & 9.85 & 0.07(2) & 0.022(5) & +0.07(11) & +0.04(2)/2 & UBV$_{pe}$, Str$_{pe}$, G \\
NGC 3228 & 7.85 & 8.35 & 0.04(1) & 0.020(4) & +0.03(10) &   & UBV$_{pe}$, Genf \\
NGC 3532 & 8.55 & 8.35 & 0.04(2) & 0.021(3) & +0.04(7) & +0.05(4)/3 & UBV$_{pe}$, Str$_{pe}$, G \\
NGC 3960 & 9.00 & 11.65 & 0.29 & 0.020(4) & +0.03(9) & +0.02(4)/1 & BV$_{pe}$, UBVI$_{ccd}$ \\
NGC 5281 & 7.90 & 10.70 & 0.22(1) & 0.019(4) & +0.00(11) &   & UBV$_{pe}$, BVI$_{ccd}$ \\
NGC 5460 & 8.20 & 9.20 & 0.13(1) & 0.021(4) & +0.06(9) & +0.05(24)/1 & UBV$_{pe}$, Str$_{pe}$, G \\
NGC 5662 & 8.05 & 9.30 & 0.30(4) & 0.022(4) & +0.06(9) &   & UBV$_{pe}$, Str$_{pe}$ \\
NGC 5999 & 8.70 & 11.75 & 0.46 & 0.019(3) & +0.00(8) &   & BV$_{pe}$, BVI$_{ccd}$ \\
NGC 6031 & 8.40 & 11.00 & 0.44(3) & 0.020(4) & +0.02(9) &   & UBV$_{pe}$, BVI$_{ccd}$ \\
NGC 6087 & 7.95 & 9.60 & 0.19(2) & 0.022(5) & +0.06(12) & +0.06(20)/1 & UBV$_{pe}$, Str$_{pe}$ \\
NGC 6134 & 9.00 & 10.00 & 0.40(3) & 0.026(5) & +0.16(9) & +0.14(2)/3 & UBV$_{pe}$, BV$_{ccd}$, Str$_{ccd}$ \\
NGC 6192 & 8.20 & 11.05 & 0.63(4) & 0.026(7) & +0.16(13) & +0.12(4)/1 & UBV$_{pe}$, BV$_{ccd}$, 2M \\
NGC 6204 & 8.05 & 10.25 & 0.47(1) & 0.021(5) & +0.05(12) &   & UBV$_{pe}$, BVI$_{ccd}$, by$_{ccd}$ \\
NGC 6281 & 8.40 & 8.65 & 0.17(3) & 0.021(3) & +0.04(7) & +0.05(6)/1 & UBV$_{pe}$, Str$_{pe}$, G \\
NGC 6405 & 7.90 & 8.45 & 0.17(2) & 0.023(5) & +0.09(11) &   & UBV$_{pe}$, Str$_{pe}$, G \\
NGC 6451 & 8.00 & 11.75 & 0.75 & 0.021(4) & +0.04(10) &   & BVI$_{ccd}$ \\
NGC 6475 & 8.40 & 7.15 & 0.07(3) & 0.024(5) & +0.11(11) & +0.09(8)/2 & UBV$_{pe}$, Str$_{pe}$, G \\
NGC 6705 & 8.40 & 11.60 & 0.39(3) & 0.027(5) & +0.18(8) & +0.17(9)/2 & UBVI$_{ccd}$ \\
NGC 6756 & 8.10 & 12.30 & 1.03(5) & 0.023(6) & +0.10(14) &   & Str$_{ccd}$ \\
NGC 6802 & 9.00 & 11.65 & 0.79 & 0.020(5) & +0.03(13) &   & BVI$_{ccd}$ \\
NGC 6830 & 8.10 & 11.90 & 0.54(4) & 0.031(3) & +0.24(5) &   & UBV$_{pe}$, $\Delta a$, 2M \\
NGC 7092 & 8.55 & 7.45 & 0.03(2) & 0.020(3) & +0.02(9) &   & UBV$_{pe}$, Str$_{pe}$, G \\
NGC 7243 & 8.00 & 9.45 & 0.24(3) & 0.021(6) & +0.06(13) &   & UBV$_{pe}$, Str$_{pe}$, G \\
NGC 7296 & 8.55 & 12.15 & 0.20 & 0.019(5) & +0.00(13) &   & BVRI$_{ccd}$ \\
Ruprecht 115 & 8.65 & 11.35 & 0.74 & 0.021(3) & +0.04(7) &   & BVI$_{ccd}$ \\
Ruprecht 130 & 8.75 & 11.50 & 1.00 & 0.020(5) & +0.03(14) &   & BVI$_{ccd}$, 2M/UKIDSS \\
Trumpler 10 & 7.80 & 8.05 & 0.03(1) & 0.016(3) & $-$0.07(10) &   & UBV$_{pe}$, APASS, Str$_{pe}$, G \\
\hline 
\end{tabular}
\tablefoot{
\tablefoottext{a}{The standard deviation of reddening is given in parentheses, if reddening was determined via photometric calibrations (e.g. $\ubv-Q$ method).}
\tablefoottext{b}{The mean spectroscopic [Fe/H] ratio and the standard deviation in parentheses. The number of studies used \citep[listed by][and additional literature]{carrera11} is given by /\#.}
\tablefoottext{c}{Photometric systems or surveys used for the analysis, e.g. G (Geneva), Str (Str\"omgren), 2M (2MASS), whereas pe/ccd stands for photoelectric and CCD data, respectively. For some clusters, $V/(b-y)$ data by \citet{mcswain05} were included (by$_{ccd}$), or $\Delta a$ photometry transformed to $V/(B-V)$ as discussed in Sect. \ref{photscale}.}
}
\end{table*}

\subsection{The results in light of lower solar abundances}
\label{newgrid}

The isochrones by \citet{poe10} are standardised on evolutionary grids with the previously accepted solar metallicity Z\,=\,0.020 \citep{anders89}. However, the solar value was meanwhile scaled down to Z\,=\,0.0134 by \citet{asplund09}. Using a reduced solar metal content, \citet{mowlavi12} present new stellar models ranging from Z\,=\,0.006 to 0.040. Unfortunately, these were provided after the beginning of our investigation, and the models are currently only available up to 3.5\,$M_{\sun}$. Nevertheless, it is worth investigating the influence of the different metallicity scales. Therefore, we searched for a suitable open cluster with sufficient photometry and well known parameters. The aforementioned mass restriction of the new models limits one to clusters with an age of $\log t \sim 8.3$, in order to be able to cover the complete main-sequence and the widest possible range in luminosity. Among our programme clusters, NGC~6475 was a good candidate for this comparison. 

\begin{figure}
\centering
\includegraphics[width=88mm]{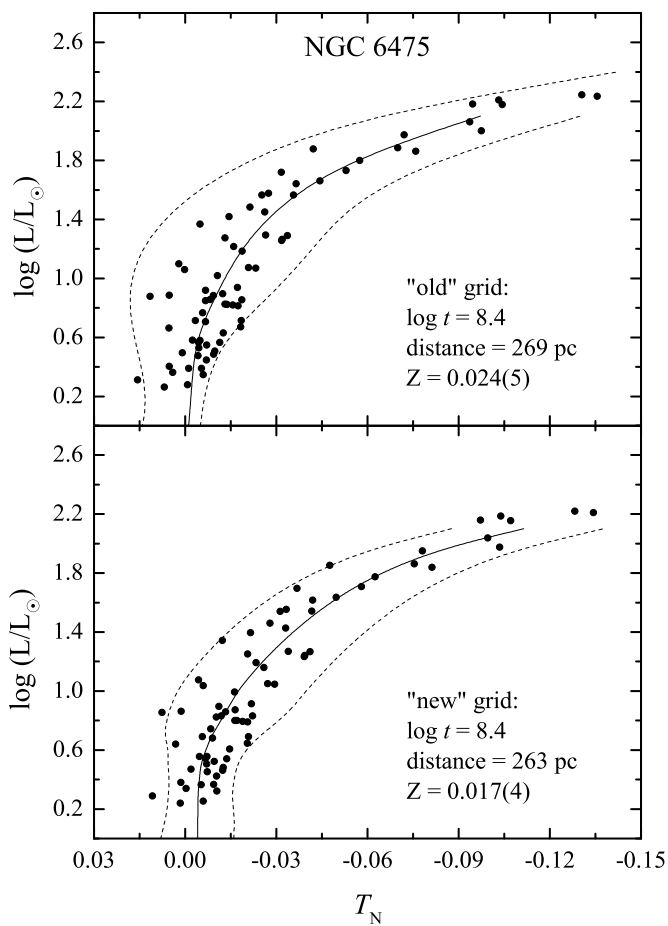}
\caption{Comparison of the old and new differential grids applied to NGC~6475. Additionally, some isochrones for the same age but for different metallicities are shown for a comparison with the dashed lines: Z=0.010/0.040 (old grid) and Z=0.010/0.030 (new grid). The effective temperature is normalised ($T_\mathrm{N}$) to the ZAMS of the respective solar metallicity.}
\label{n6475comp}
\end{figure}

Following \citet{poe10}, we constructed new differential grids based on the ones by \citet{mowlavi12} with their Z\,=\,0.014 model as a reference. This was performed for ages around our final result of $\log t=8.4$, which is in very good agreement with the starting value ($\log t=8.34$) compiled from the literature. The best fit (Fig. \ref{n6475comp}) was obtained with the same age ($\log t=8.4$) and Z\,=\,0.017(4). Using equation A.5 by \citet{mowlavi12} this resulted in a metallicity of [Fe/H]\,=\,0.14\,dex, whereas the original grids gave a slightly lower value of [Fe/H]\,=\,0.11\,dex derived using Z\,=\,0.024(5). The applied distances agree to within a few percent; 263\,pc (new grid) vs. 269\,pc (old grid). Both of them are very close to the Hipparcos based result of 270\,pc \citep{leeuwen09}. One can conclude that the DG method is independent of the grids used. The resulting Z values only need to be transformed into the corresponding [Fe/H] values. Our results are in good agreement with the mean spectroscopic iron abundance of 0.09(8)\,dex, which was derived from two sources: [Fe/H]\,=\,0.14(6)\,dex by \citet{sestito03} and 0.03(2)\,dex by \citet{villanova09}. Differing metallicities are also found for other open clusters (e.g. Melotte~20 or NGC~2632). By using the spectral resolution and the number of investigated stars as criteria to determine the most reliable result, the higher value by \citet{sestito03} should be considered.   

\section{Comparison with spectroscopic results}
\label{speccomp}
\citet{carrera11} present new abundance determinations for a number of open clusters, as well as a comprehensive list of high-resolution spectroscopic [Fe/H] results compiled from the literature. We used this list as a reference, along with studies by \citet{an07}, \citet{pancino10}, \citet{schuler10}, and \citet{santos12}.  
All measurements were based on spectroscopic data with a resolution of at least R\,=\,16\,000, however the majority were obtained with R\,=\,40\,000 or higher. Where several studies were available for one cluster, we calculated a mean [Fe/H] value and its standard deviation. Otherwise, we adopted the individual results and quoted errors from these references. 

\begin{figure}
\centering
\includegraphics[width=88mm]{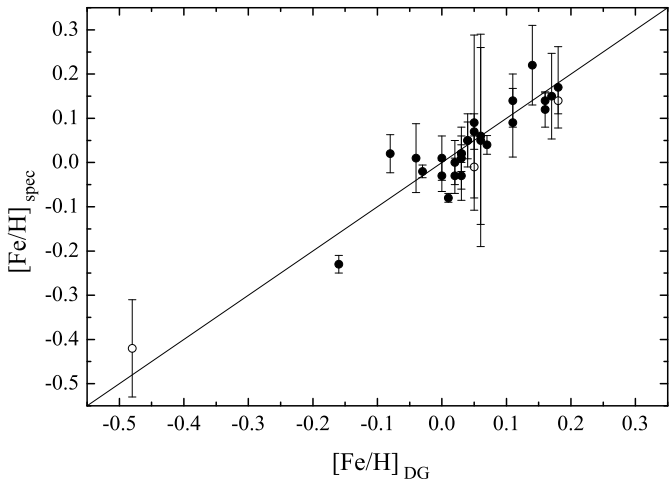}
\caption{Comparison of our metallicity determinations to high-resolution spectroscopic values from the literature. Open symbols are additional open clusters by \citet{poe10}, and the straight line represents the one-to-one relation. The errors of our results are about 0.1\,dex for most clusters, but were omitted for better visibility.}
\label{gridcomp}
\end{figure}

For 27 programme open clusters, spectroscopic determinations were found in the literature. Furthermore, we included the DG results for three open clusters (Melotte~25, NGC~752, and Berkeley~29) investigated by \citet{poe10}, due to the availability of spectroscopic data for them. The last cluster is of particular interest, because it is the most underabundant aggregate in the complete sample so far. This allowed for a better verification of the metallicity scale obtained by the DG method (see Fig. \ref{gridcomp}). 

The largest deviation (0.15\,dex) between spectroscopic results and our determination was initially found for IC~4725. The iron abundance [Fe/H]\,=\,0.18(8)\,dex was derived by \citet{luck94}, as listed by \citet{carrera11}. However, for the three investigated cluster stars, \citet{luck00} present revised metallicities based on spectra with much higher resolution (R\,=\,60\,000 compared to R\,=\,18\,000). The resulting mean [Fe/H]\,=\,0.02(2)\,dex is in excellent agreement with our study ([Fe/H]\,=\,0.03(8)\,dex). Somewhat less deviating results were found for the open clusters Melotte~22 (Pleiades) and NGC~1901 with differences of 0.10 and 0.09\,dex, respectively. The spectroscopic determination for NGC~1901 was based on a single star analysed by \citet{carraro07}. The discrepancy for the prominent and well investigated Pleiades cluster could be due to the transformation of the overall metallicity Z to [Fe/H], assuming a correlation between the various abundance ratios. The analysis by \citet{geb08} showed a deficiency in the main contributors to Z: the elements C and O. Therefore, our derived iron abundance of [Fe/H]\,=\,$-$0.08(7)\,dex for the Pleiades is also very probably underestimated. 

Using the complete sample of 30 open clusters with available spectroscopic metallicity determinations, and adopting the revised result for IC~4725, we concluded that the DG method matches the spectroscopic scale very well (Fig. \ref{gridcomp}). The mean value of the absolute deviations amounts to 0.03(3)\,dex, without a  noticeable offset or gradient. However, more aggregates in the underabundant regime are still necessary for a more detailed comparison.

\section{Conclusion}

We examined 58 open clusters in total using the differential grid method introduced by \citet{poe10}. The comparison of the derived cluster metallicities with spectroscopic metallicity determinations showed that accurate results could be achieved. Although the mean deviations to spectroscopic studies were lower than 0.05\,dex for [Fe/H], the comparison with results by \citet{poe10} showed that improperly applied distances can lead to errors that are somewhat larger ($\sim$\,0.1 dex). However, differences of that order are also not uncommon among spectroscopic studies. Examining the compilation by \citet{carrera11}, we found results with much greater discrepancies (e.g. for Melotte~20, NGC~2632, or Collinder~261).

To verify the effect of the currently accepted solar abundance value \citep{asplund09}, the result for NGC~6475 using the original differential grids by \citet{poe10} were compared with newly constructed grids based on the recent evolutionary models by \citet{mowlavi12}. Since the derived parameters were in excellent agreement, we concluded that the influence of a different solar metallicity (Z\,=\,0.02 compared to Z\,=\,0.014) is negligible. The resulting metallicity values only need to be properly transformed into [Fe/H] values. However, more comparisons will be needed as soon as the new models are extended to higher masses.

Nevertheless, this method provides a robust way of obtaining accurate metallicity estimates for large samples of open clusters, necessary for a more in-depth study of various relationships, such as metallicity gradients or age dependencies. The future availability of deep and homogeneous photometric surveys (e.g. LSST, Pan-STARRS, Vista) will allow the study of cluster metallicities in the very inner and outer regions of the Milky Way in a consistent way. Spectroscopic investigations, on the other hand, are very time-consuming, therefore only about 100 open clusters are covered so far. Furthermore, for most targets, the overall cluster metallicity has been defined by three or less stars (as a rule red giants, the brightest objects). The influence of improper membership determination in these number limited samples must also be considered. 

As soon as accurate astrophysical parameters are known for numerous individual cluster stars, the DG method allows evolutionary models to be tested. In this respect we would like to mention the upcoming Gaia satellite mission, which will provide spectroscopic effective temperatures and metallicities (limited in distance and temperature domain), as well as parallax, radial velocity, and proper motion data, to identify true open cluster members and to derive luminosity. With these data sets at hand, our method can be extended to investigate topics related with rotation. This parameter was recently implemented in the models by the Geneva group \citep{ekstroem12}. Furthermore, an extension of the differential grids to post main-sequence stages will be helpful when investigating older stellar aggregates (e.g. globular clusters) in more detail.

The investigated target sample comes from the $\Delta a$ photometric survey \citep[e.g.][]{net07}, which aims to detect chemically peculiar stars in open clusters. In a follow-up study, the results of the present paper will be used to investigate the dependencies of this star group with age and metallicity.

\begin{acknowledgements}
This publication makes use of data products from the Two Micron All Sky Survey, which is a joint project of the University of Massachusetts and the Infrared Processing and Analysis Center/California Institute of Technology, funded by the National Aeronautics and Space Administration and the National Science Foundation, and was made possible through the use of the AAVSO Photometric All-Sky Survey (APASS), funded by the Robert Martin Ayers Sciences Fund. 
Furthermore, this research has made use of the WEBDA database, operated
at the Department of Theoretical Physics and Astrophysics of the Masaryk University. 
This work was supported by the grants GA \v{C}R P209/12/0217, 7AMB12AT003, 
the Austrian Research Fund via the project FWF P22691-N16,
and the financial contributions of the Austrian Agency for International
Cooperation in Education and Research (CZ-10/2012). We would like to thank Kieran Leschinski for his help with language editing.
\end{acknowledgements}

\Online

\begin{figure*}
\centering
\includegraphics[width=15cm]{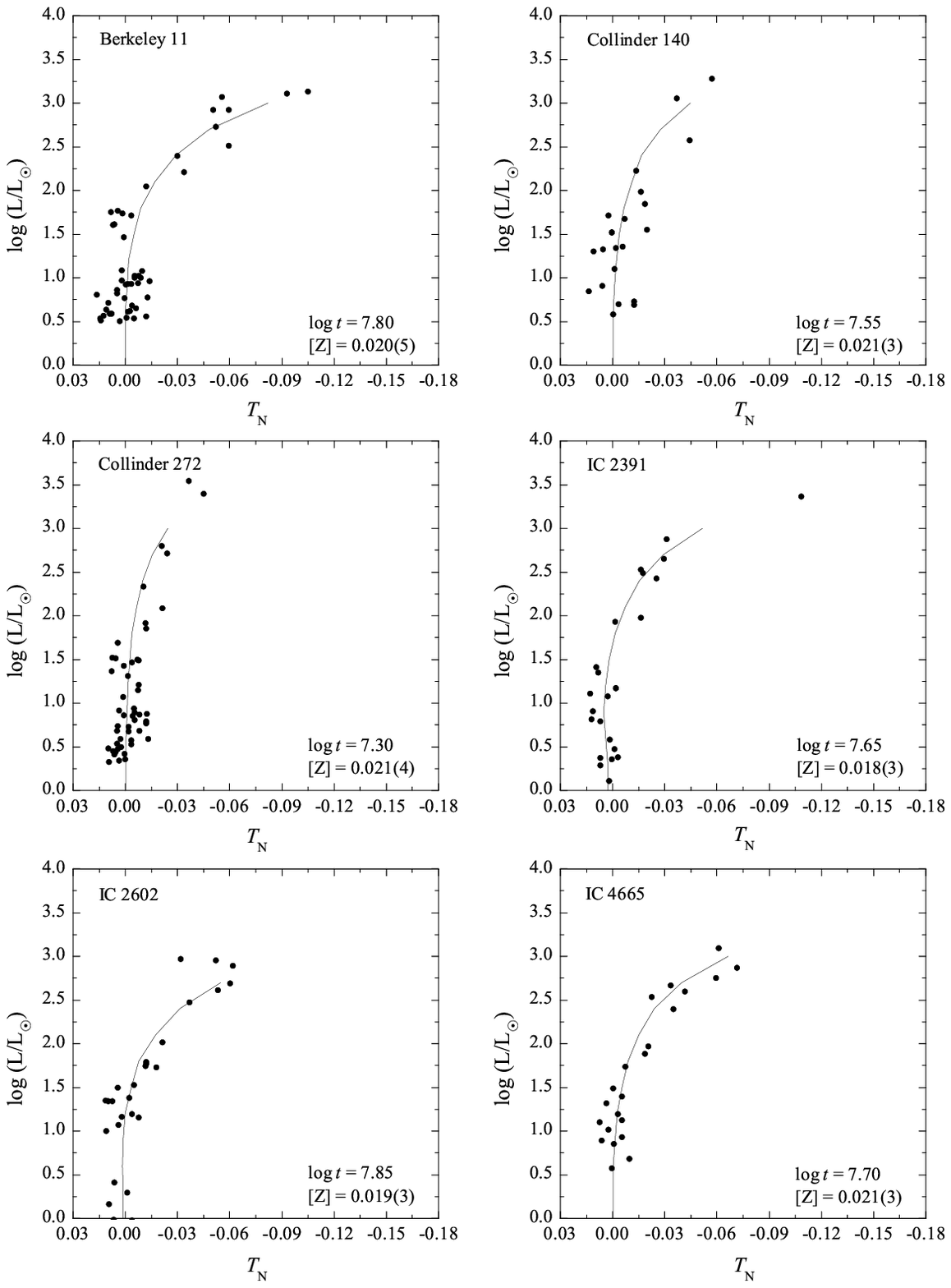}
\caption {The best fits of the differential grids by \citet{poe10} to data from the programme open clusters, in order to determine the fundamental parameters including metallicity. For clarity, $T_{N}$ are effective temperatures normalised to the ZAMS. An equal scale is used for a better comparison of different ages and metallicities. Owing to interpolation in the available grids, the most luminous stars are sometimes not covered by the isochrone; however, for older clusters we also included evolved stars to illustrate the continuation of the cluster sequences.} \label{plot1}
\end{figure*}

\begin{figure*}
\centering
\includegraphics[width=155mm]{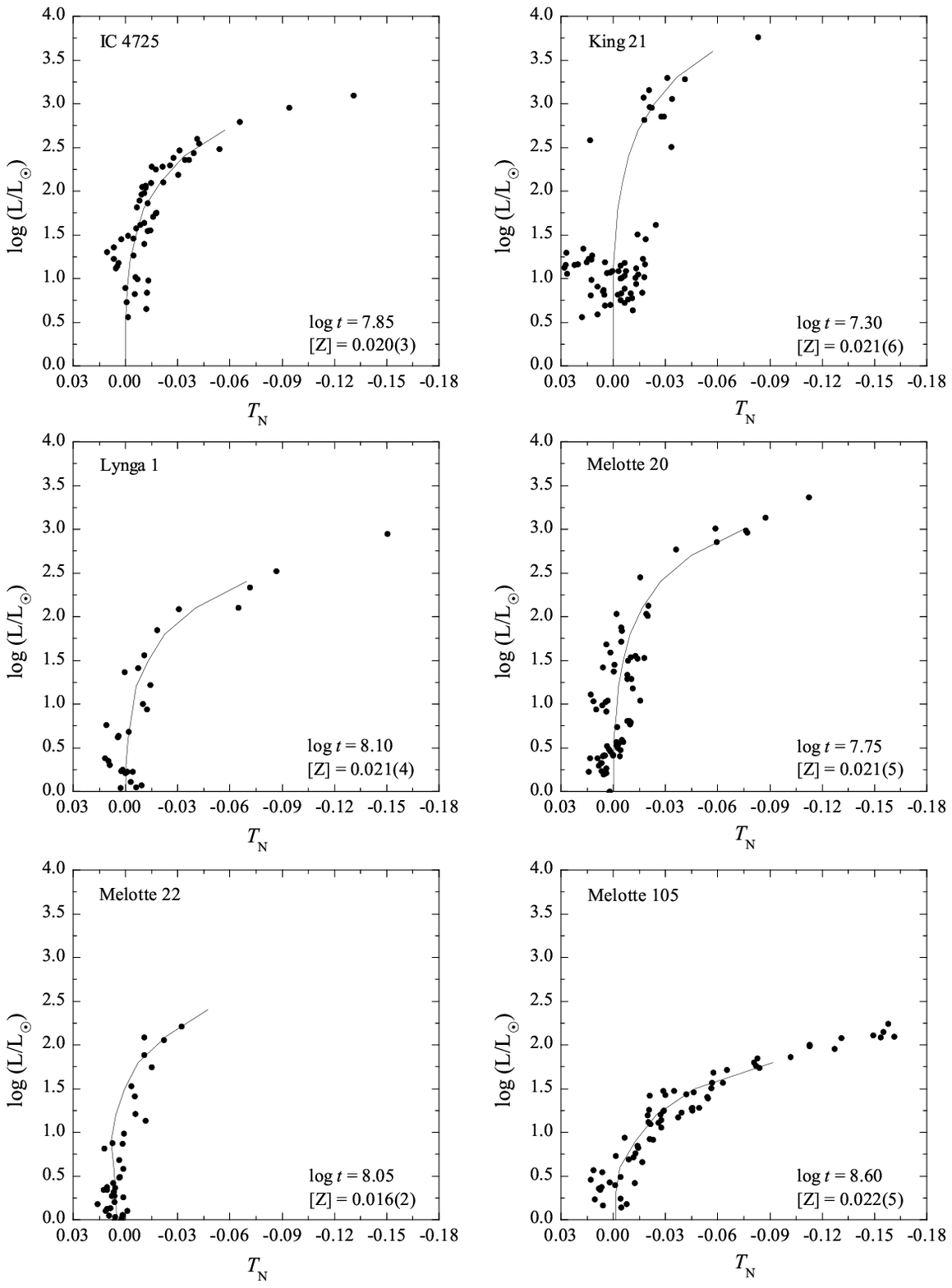}
\caption {Continuation of Figure \ref{plot1}.} \label{plot2}
\end{figure*}

\begin{figure*}
\centering
\includegraphics[width=155mm]{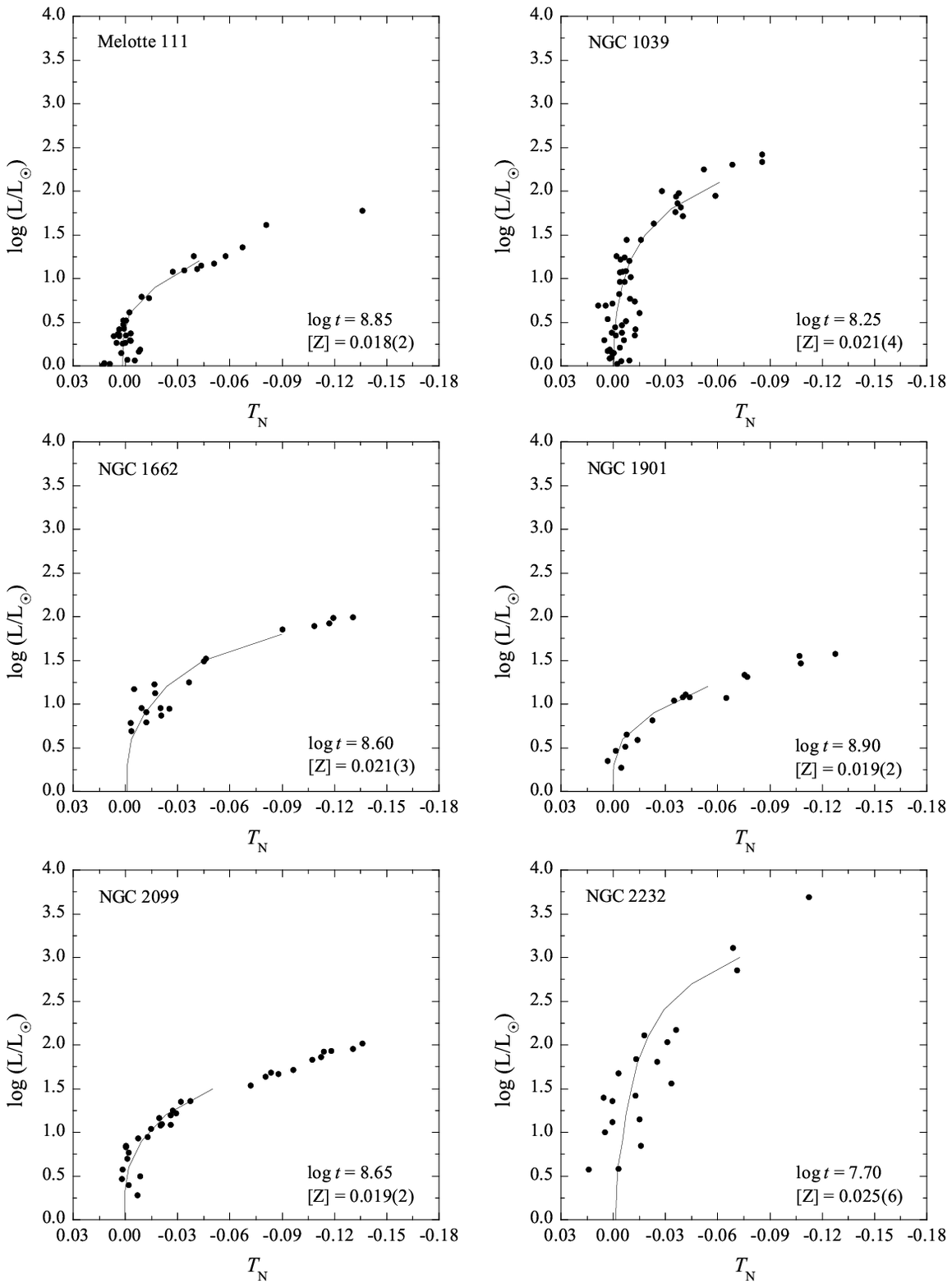}
\caption {Continuation of Figure \ref{plot1}.} \label{plot3}
\end{figure*}

\begin{figure*}
\centering
\includegraphics[width=155mm]{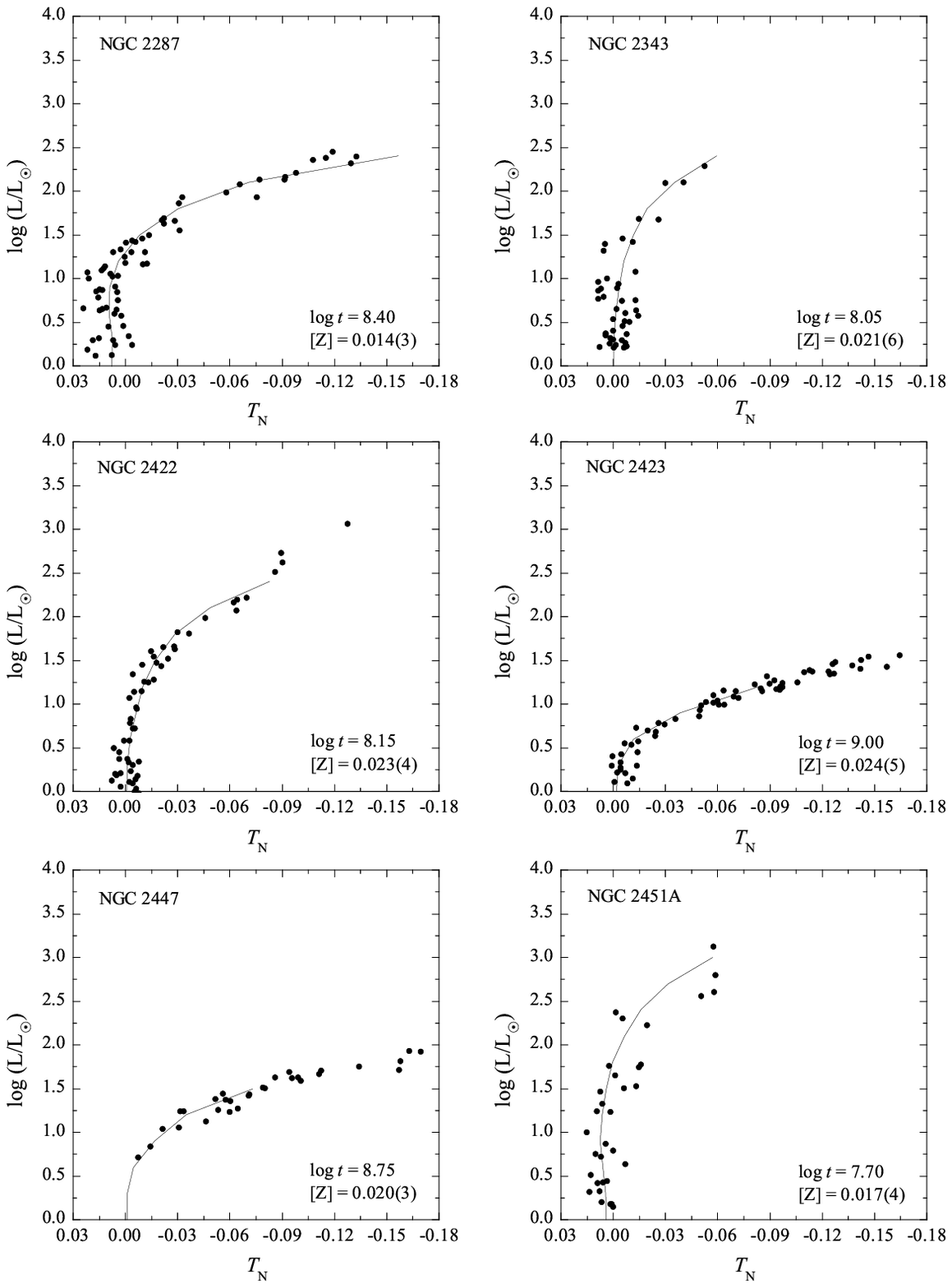}
\caption {Continuation of Figure \ref{plot1}.} \label{plot4}
\end{figure*}

\begin{figure*}
\centering
\includegraphics[width=155mm]{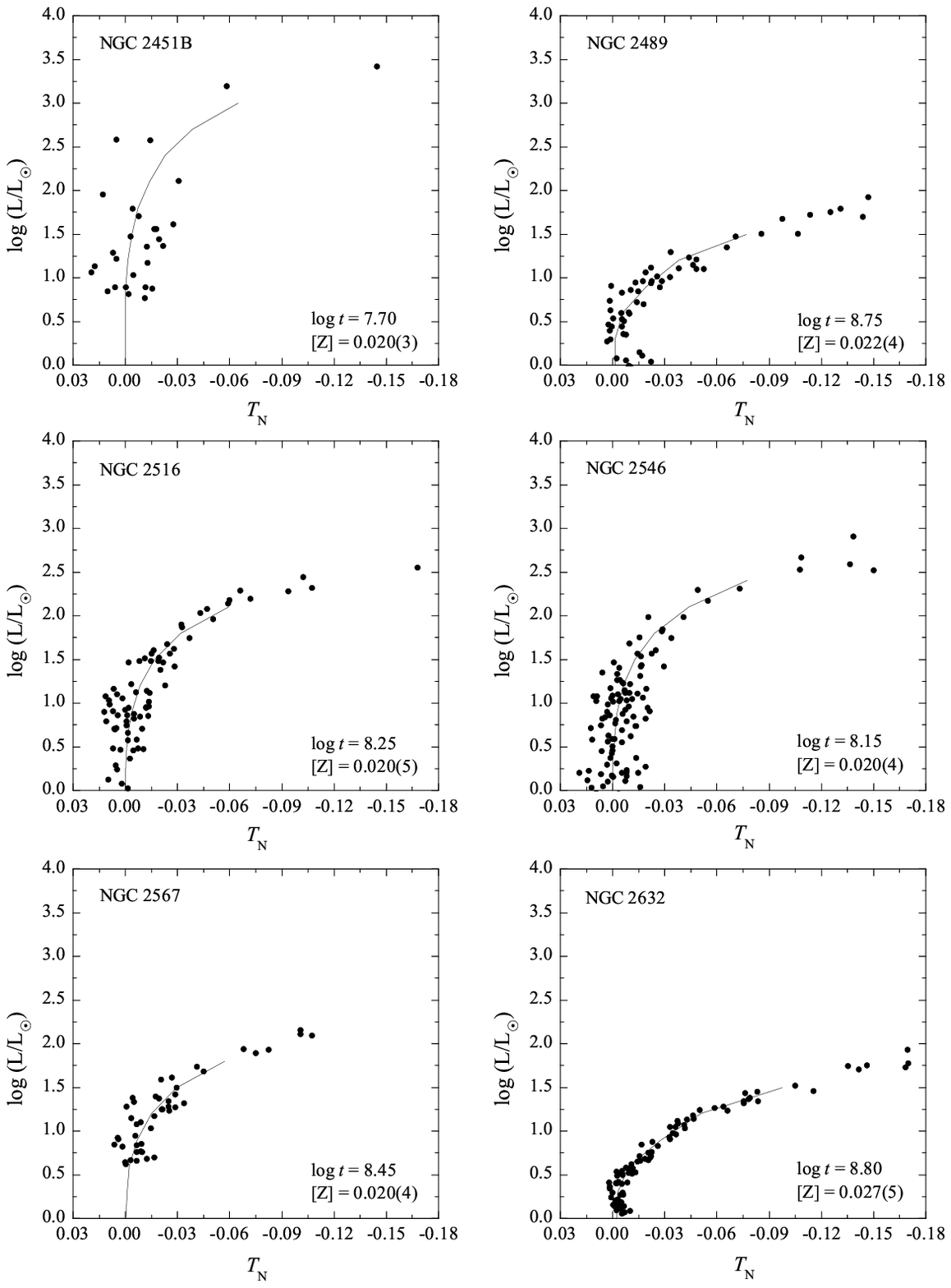}
\caption {Continuation of Figure \ref{plot1}.} \label{plot5}
\end{figure*}

\begin{figure*}
\centering
\includegraphics[width=155mm]{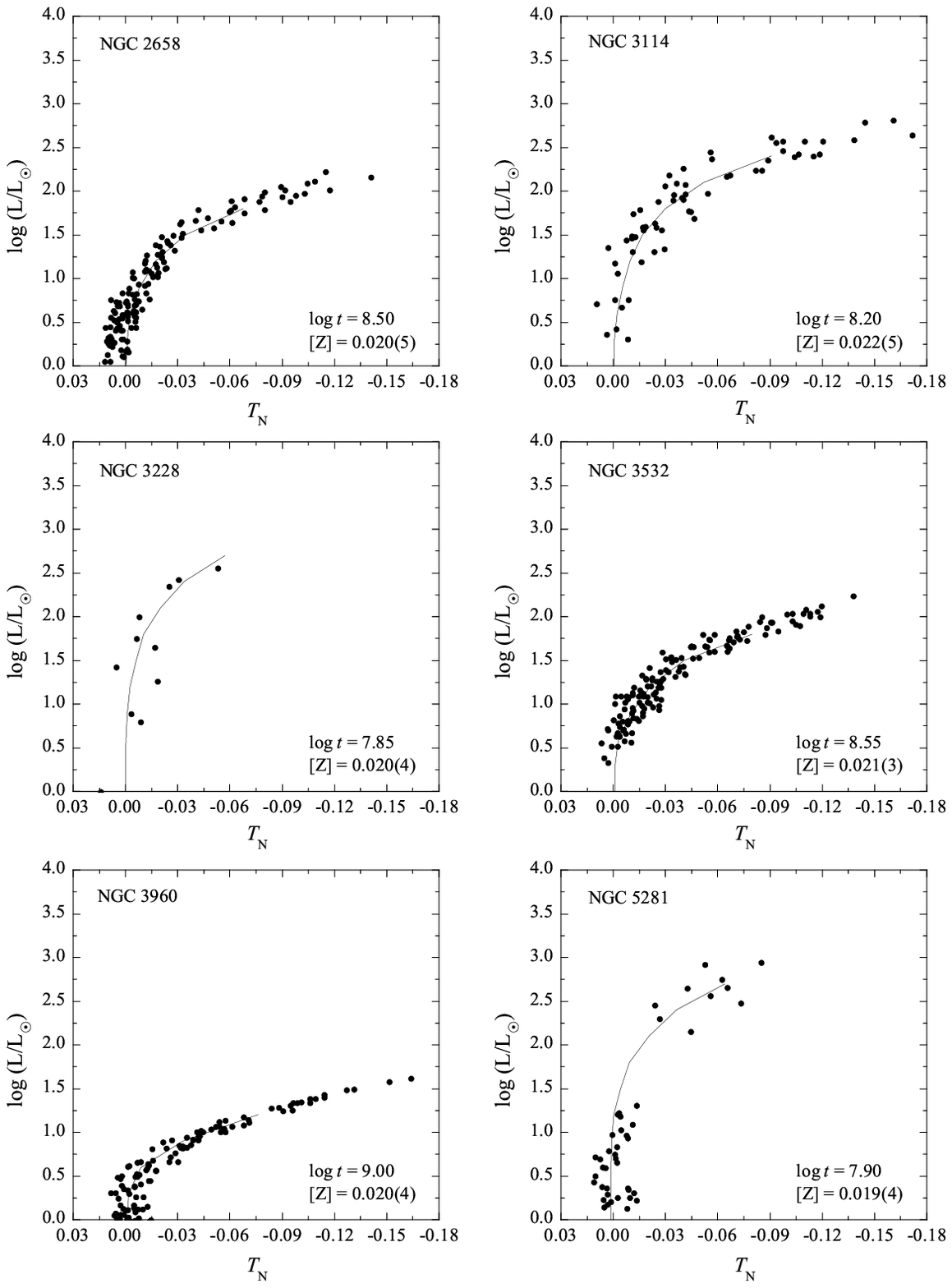}
\caption {Continuation of Figure \ref{plot1}.} \label{plot6}
\end{figure*}

\begin{figure*}
\centering
\includegraphics[width=155mm]{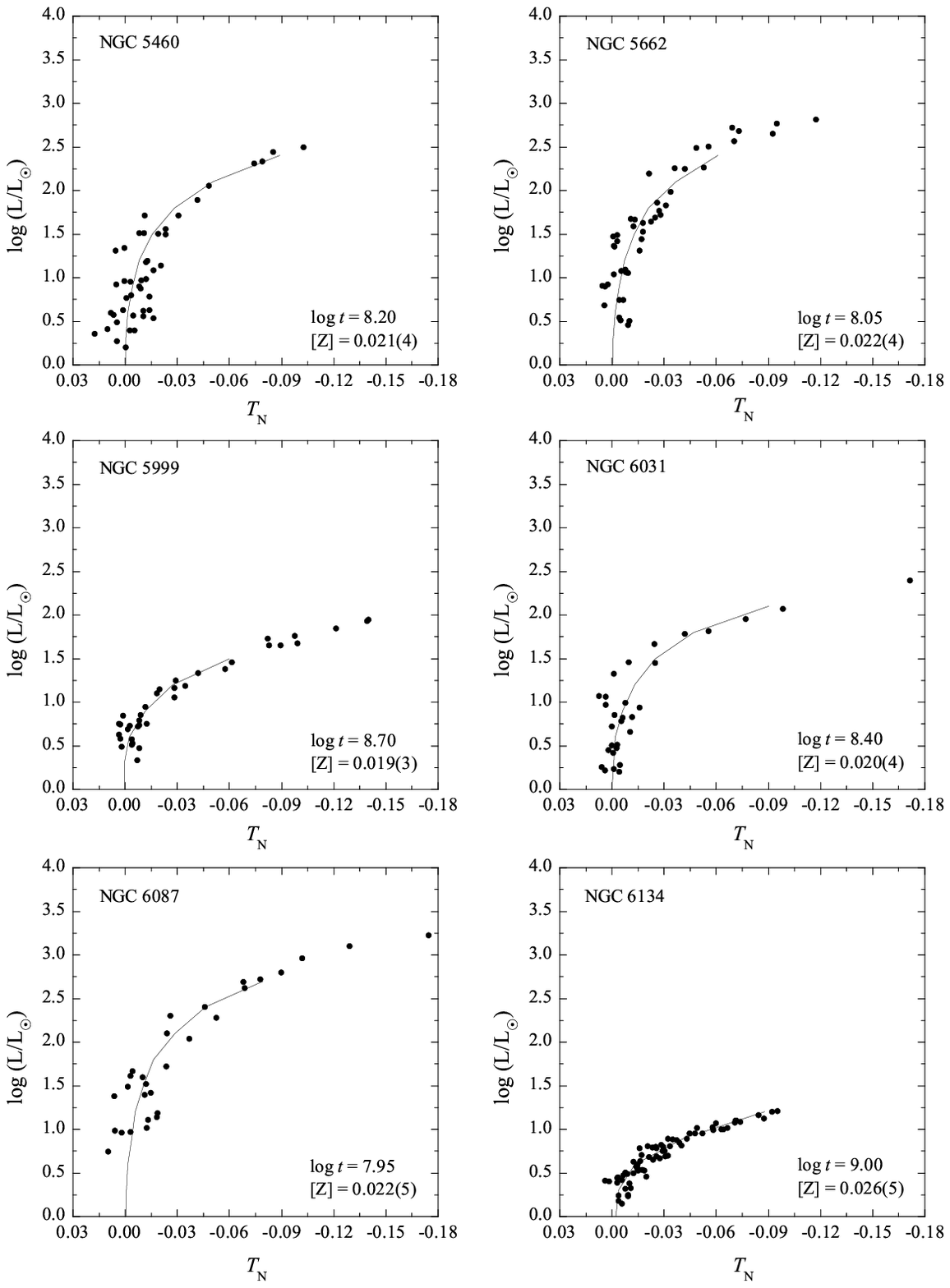}
\caption {Continuation of Figure \ref{plot1}.} \label{plot7}
\end{figure*}

\begin{figure*}
\centering
\includegraphics[width=155mm]{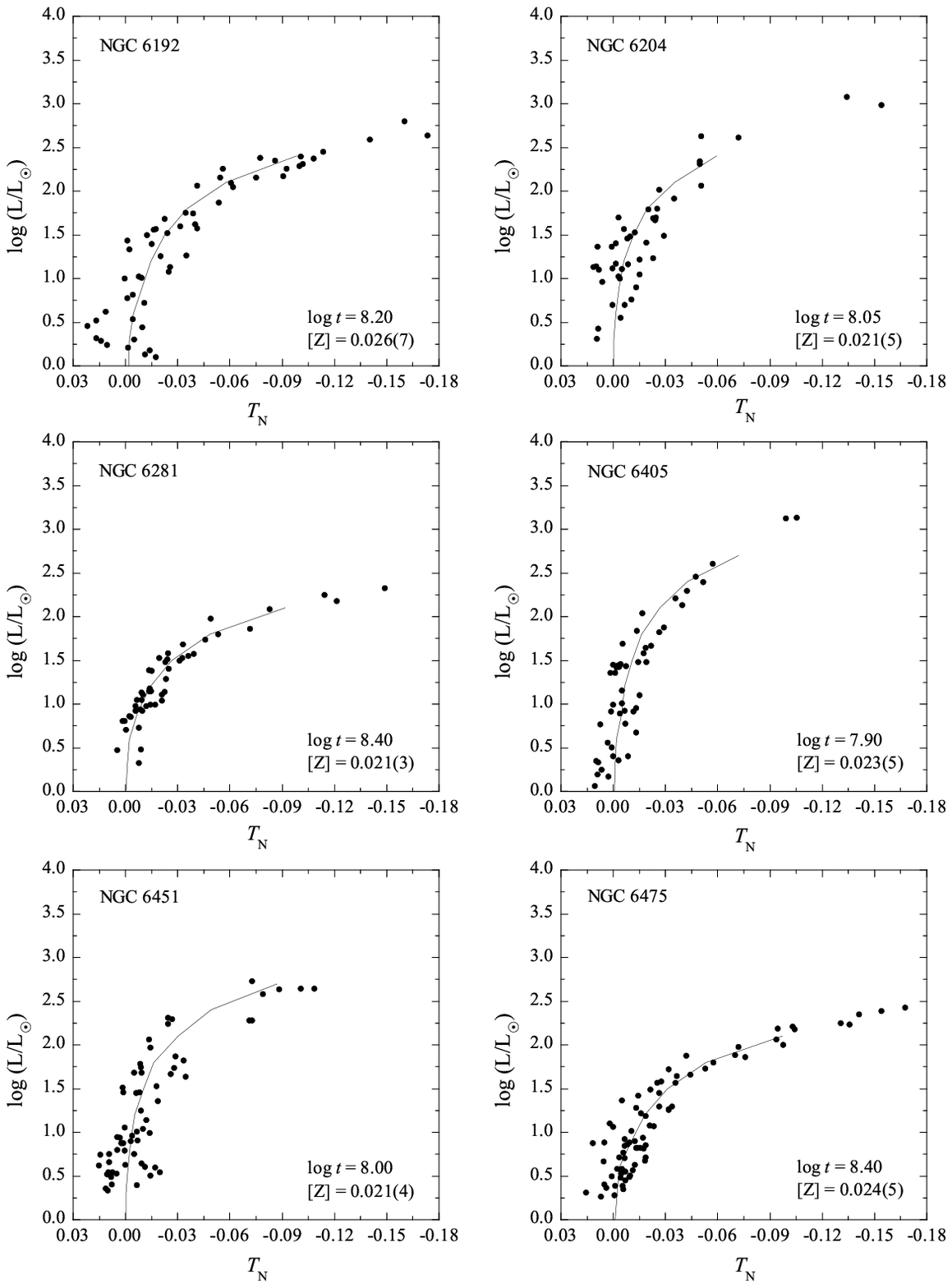}
\caption {Continuation of Figure \ref{plot1}.} \label{plot8}
\end{figure*}

\begin{figure*}
\centering
\includegraphics[width=155mm]{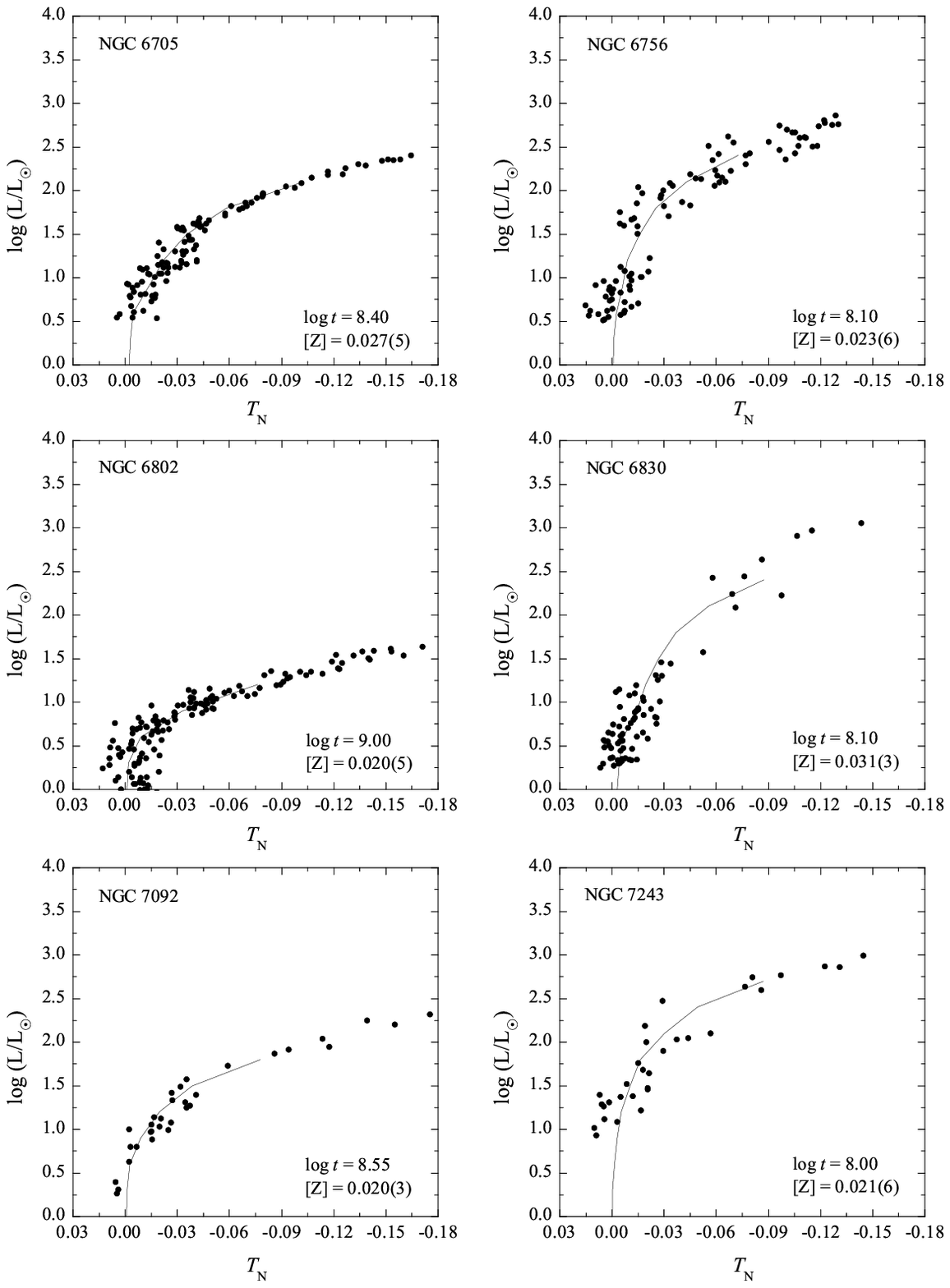}
\caption {Continuation of Figure \ref{plot1}.} \label{plot9}
\end{figure*}

\begin{figure*}
\centering
\includegraphics[width=155mm]{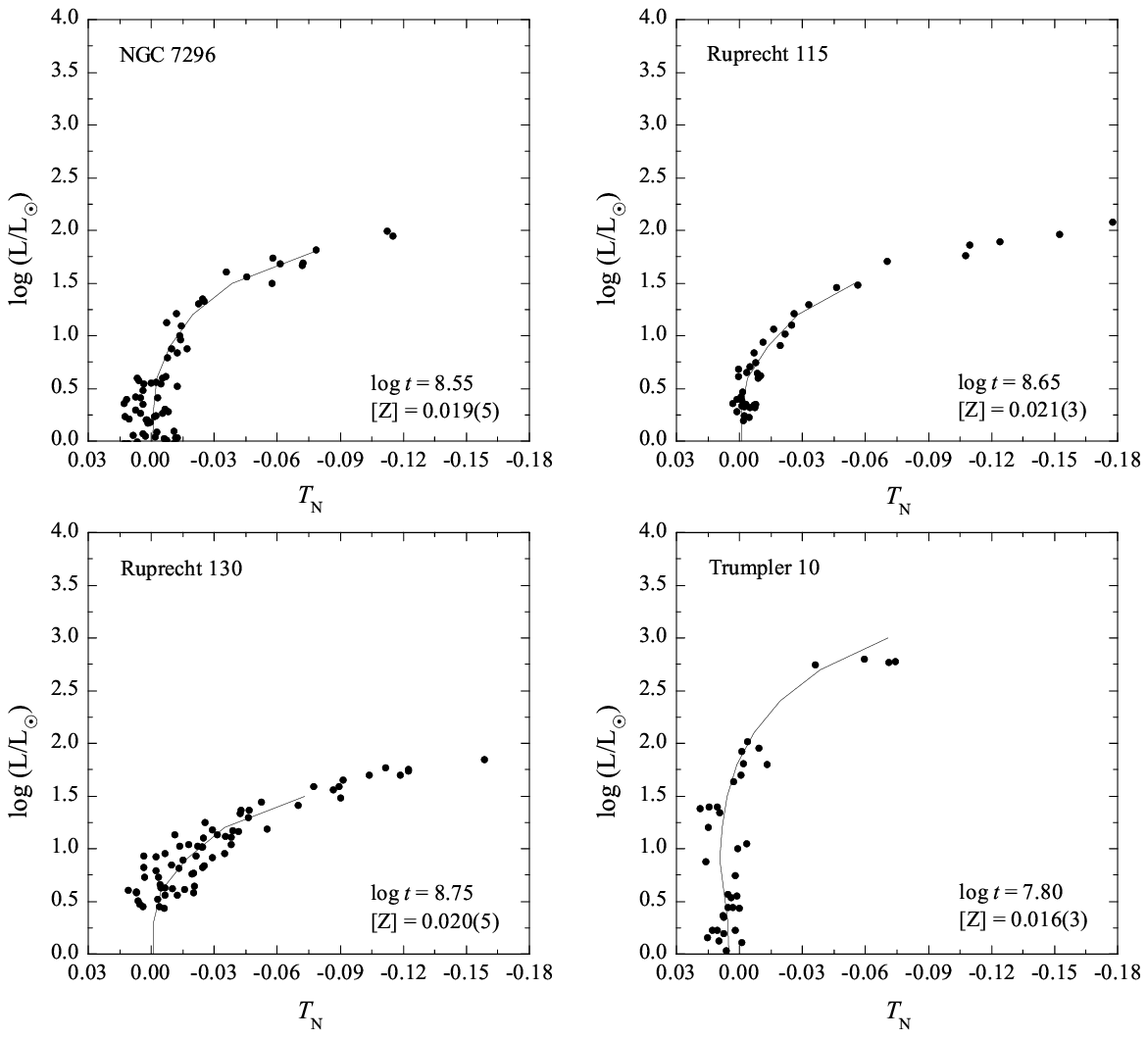}
\caption {Continuation of Figure \ref{plot1}.} \label{plot10}
\end{figure*}

\end{document}